\pdfoutput=1

\PassOptionsToPackage{dvipsnames}{xcolor}
\documentclass[11pt]{article}

\usepackage[final]{acl}

\usepackage{times}
\usepackage{latexsym}

\usepackage[T1]{fontenc}

\usepackage[utf8]{inputenc}

\usepackage{microtype}

\usepackage{inconsolata}

\usepackage{graphicx}

\usepackage{booktabs}
\usepackage{makecell}
\usepackage{graphicx} 
\usepackage{array} 
\usepackage{arydshln}
\usepackage{multirow}
\usepackage{tabularray}
\usepackage{adjustbox}
\usepackage{placeins}

\newcolumntype{L}[1]{>{\raggedright\let\newline\\\arraybackslash\hspace{0pt}}m{#1}}
\newcolumntype{C}[1]{>{\centering\let\newline\\\arraybackslash\hspace{0pt}}m{#1}}
\newcolumntype{R}[1]{>{\raggedleft\let\newline\\\arraybackslash\hspace{0pt}}m{#1}}

\makeatletter
\def\adl@drawiv#1#2#3{%
        \hskip.5\tabcolsep
        \xleaders#3{#2.5\@tempdimb #1{1}#2.5\@tempdimb}%
                #2\z@ plus1fil minus1fil\relax
        \hskip.5\tabcolsep}
\newcommand{\cdashlinelr}[1]{%
  \noalign{\vskip\aboverulesep
           \global\let\@dashdrawstore\adl@draw
           \global\let\adl@draw\adl@drawiv}
  \cdashline{#1}
  \noalign{\global\let\adl@draw\@dashdrawstore
           \vskip\belowrulesep}}
\makeatother

\definecolor{purple}{RGB}{100, 0, 200}

\usepackage{import}
\usepackage{graphicx}
\usepackage{amsmath}
\usepackage{amssymb}
\usepackage{booktabs}
\usepackage{float}
\usepackage{listings}
\usepackage{multirow}
\usepackage{lipsum}  
\usepackage{subcaption}
\usepackage{comment}
\usepackage[switch]{lineno}
\usepackage{wrapfig}
\usepackage{varwidth}
\usepackage{array}
\usepackage{makecell}
\usepackage[capitalize]{cleveref}
\usepackage{todonotes}
\usepackage[normalem]{ulem}
\useunder{\uline}{\ul}{}

\usepackage{pifont}

\newcommand{\rinlinecode}[1]{\lstinline[basicstyle=\ttfamily, breaklines=true, breakatwhitespace=true]|#1|}

\newcommand{\infoncesynth}{InfoNCE \textsubscript{Synth.}}
\newcommand{\wssynth}{WS \textsubscript{Synth.}}
\newcommand{\infoncereal}{InfoNCE \textsubscript{Real}}
\newcommand{\wswithreal}{WS \textsubscript{Synth. + Real}}
\newcommand{\hab}{\hphantom{$^{ab}$}}
\newcommand{\vab}{$^{ab}$}
\newcommand{\vaa}{$^{a\hphantom{b}}$}
\newcommand{\vbb}{$^{b\hphantom{a}}$}

\title{Beyond Contrastive Learning: Synthetic Data Enables List-wise Training with Multiple Levels of Relevance}

\author{
\textbf{Reza Esfandiarpoor\thanks{\footnotesize{Equal contributions.}}}$^{1}$ \quad
\textbf{George Zerveas}$^{*2}$ \quad
\textbf{Ruochen Zhang}$^{1}$ \quad
\textbf{Macton Mgonzo}$^{1}$ \quad
\\
\textbf{Carsten Eickhoff}$^{\,3}$
\textbf{Stephen H. Bach}$^{1}$
\\
$^{1}$Brown University \quad
$^{2}$Microsoft \quad
$^{3}$University of T\"{u}bingen \\
\texttt{\{reza\_esfandiarpoor,ruochen\_zhang,macton\_mgonzo,stephen\_bach\}@brown.edu} \\
\texttt{gzerveas@microsoft.com} \quad
\texttt{c.eickhoff@acm.org} \\
}

\begin{document}
\maketitle

\begin{abstract}
Although synthetic data has changed various aspects of information retrieval (IR) pipelines, the main training paradigm remains: contrastive learning with binary relevance labels, where one positive document is compared against several negatives using the InfoNCE loss.
This objective treats all documents that are not explicitly annotated as relevant on an equally negative footing, regardless of their actual degree of relevance, thus missing subtle nuances useful for ranking.
To overcome this limitation, in this work, we forgo real documents and annotations and use large language models to directly generate synthetic documents that answer the MS~MARCO queries according to \emph{several different levels of relevance}.
We also propose using Wasserstein distance as a more effective loss function for training transformer-based retrievers with graduated relevance labels.
Our experiments on MS~MARCO and BEIR benchmark show that our proposed approach outperforms conventional training with InfoNCE by a large margin. 
Without using any real documents, our method significantly improves self-supervised retrievers and is more robust to distribution shift compared to contrastive learning using real data.
Our method also successfully integrates existing real data into the synthetic ranking context, further boosting the performance.
Overall, we show that generating multi-level ranking contexts is a better approach to synthetic data generation for IR than just generating the standard positive and negative documents.
Code: \url{https://github.com/BatsResearch/sycl}
\end{abstract}

\begin{figure}[t!]
  \centering
  \includegraphics{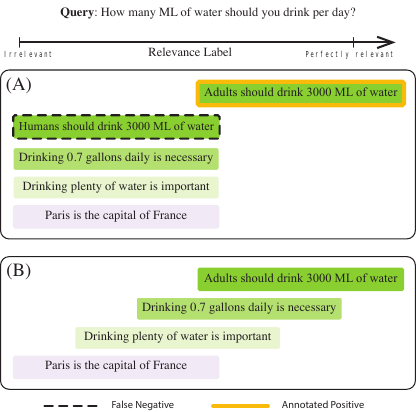}
  \caption{A) Standard contrastive training with real data treats all passages except the explicitly annotated positive passage the same, on a binary basis, regardless of their actual similarity to the given query. It is also vulnerable to false negatives. B) SyCL generates a synthetic multi-level ranking context and trains the model to rank passages based on their degree of relevance to the given query.}
  \label{fig:intro_fig}
\end{figure}
\section{Introduction}
\label{sec:intro}

The ability of information retrieval (IR) methods to rank a collection of documents based on their relevance to a given query is critical for many applications like web search and, more recently, retrieval augmented generation (RAG)~\citep{lewis2020retrieval,shi2023replug}.
However, since existing large-scale IR datasets only provide binary relevance labels~\citep{bajaj_msmarco_2018}, most recent work predominantly trains retrievers to simply separate relevant from irrelevant documents~\citep{ma2025drama}.
This implicitly assumes that the similarity metric learned through this simple training objective is precise enough during inference to rank multiple relevant documents differing in very nuanced ways.
Instead, here we use large language models (LLMs) to generate multiple synthetic documents with \textit{graduated relevance levels} for each query, which enables us to explicitly guide retrievers to \textit{rank a collection of documents} during training.

Most large-scale IR datasets only provide binary relevance labels that divide documents into relevant (``positive'') and irrelevant (``negative'') (\cref{fig:intro_fig}A).
Moreover, they contain very few, often only one, positive(s) per query~\citep{bajaj_msmarco_2018}.
Similarly, even the most recent synthetic datasets adopt a binary definition of relevance~\citep{weller2024promptriever}.
These limitations are reflected in the predominant training paradigm: contrastive learning with the InfoNCE loss~\citep{oord2019representationlearningcontrastivepredictive}.
However, this objective differs from ranking in that all documents other than a single annotated positive are treated as negatives of equal non-relevance, regardless of their actual semantic similarity to the query.
Additionally, it only considers a single relevant document at each training step.
By contrast, an effective retriever is expected to \textit{rank a collection of documents} with potentially multiple positives according to nuanced semantic differences.
Also, since existing datasets like  MS~MARCO are sparsely annotated, many unannotated positives are falsely used as negatives, which further degrades the training signal~\citep{qu-2021-rocketqa}.

On the other hand, the benefits of a ranking context (i.e., annotated documents for each query) with multiple relevance levels are well established in the \textit{learning-to-rank} (L2R) literature~\citep{cao2007learning,ai_learning_2019, ai_learning_2018}.
Most L2R works date before the advent of transformers and rely on small datasets with engineered features~\citep{qin_letor_2010, pmlr-v14-chapelle11a,dato_fast_2016}, which are not useful for training contemporary retrievers.
To train transformer-based retrievers with fine-grained annotations, some have used cross-encoders to pseudo-label the top retrieved documents for each query~\citep{wang2022text}.
However, because of the sparse annotations, the candidate documents are often mostly unannotated positives.
In general, it is challenging to select a small set of candidate documents that covers a wide range of relevance levels, i.e., from irrelevant to perfectly relevant
(see \cref{sec:apndx_pseudo_lbl} for a detailed discussion).
Besides, pseudo-labeling is not applicable where existing data is scarce, such as niche domains like climate research or new tasks like retrieval with reasoning or instructions~\citep{su2024bright,weller2024promptriever}.
As an alternative, LLMs provide a unique opportunity to generate rich ranking contexts without these limitations.

In this paper, we propose  SyCL (\textbf{Sy}nthetic ranking \textbf{C}ontext for \textbf{L}ist-wise training), a novel approach that enables training large transformer-based retrievers with graduated relevance labels.
First, we create a large-scale IR dataset (\textasciitilde 2M passages) that provides several passages with different relevance levels for each query (\cref{fig:intro_fig}B).
To avoid data annotation problems (e.g., sparsity and noise) while maintaining diversity and scale, we forgo real documents and use LLMs to generate synthetic documents with four different relevance levels for training queries of the MS MARCO dataset.
During training, our dataset allows us to penalize the model's scoring choices differently depending on the relative degree of disagreement between predicted and ground-truth relevance.
Second, we propose to use the Wasserstein distance as a list-wise loss function that can effectively leverage graduated relevance labels to optimize large transformer-based retrievers. 

Through extensive experiments, we show the importance and effectiveness of multi-level ranking contexts.
Without using any real documents, SyCL significantly improves the performance of self-supervised retrievers in both in-domain evaluation on MS MARCO and zero-shot evaluation on BEIR~\citep{thakur2021beir}.
Most importantly, we show that generating multi-level ranking contexts instead of just positives and negatives is a better approach to synthetic data generation for IR.
Specifically, training on graduated relevance labels improves the nDCG@10 score compared to training on the same synthetic data with binary labels by 5.5 and 6.4 points on average for MS~MARCO and BEIR respectively.
Using synthetic data alone, SyCL outperforms training on binary \textit{real data} on out-of-domain evaluation on BEIR by an average of 2.3 nDCG@10 points.
Moreover, we successfully integrate real data into the synthetic ranking context, which achieves better performance than both synthetic and real data alone.
Through additional analytical experiments, we show the individual significance of the Wasserstein loss and graduated relevance labels.
Finally, we analyze our data generation pipeline and find that even small 32B LLMs can generate high-quality training data.
We summarize our main contributions as follows:
\begin{itemize}
    \item We introduce SyCL, a novel method for training dense retrievers, which (a) uses publicly available LLMs to generate a large corpus of synthetic documents with graduated relevance labels and (b) uses Wasserstein distance as a list-wise loss function for training with multiple relevance levels.
    
    \item We show that using the same synthetic data, training with multiple levels of relevance outperforms standard contrastive training with binary relevance labels and InfoNCE loss.
    
    \item We show that without using real documents, SyCL significantly boosts the performance of self-supervised retrievers and is more robust to distribution shift, outperforming contrastive learning with real binary data in zero-shot evaluation on BEIR.
    SyCL can also combine real and synthetic data to further boost performance.

\end{itemize}

Our work uncovers the potential of LLMs for generating datasets that offer a more fine-grained definition of relevance compared to existing training data.
Our findings encourage future work to explore novel data generation methods that better represent the retrieval task.
\begin{figure*}[t!]
  \centering
  \includegraphics[width=\textwidth]{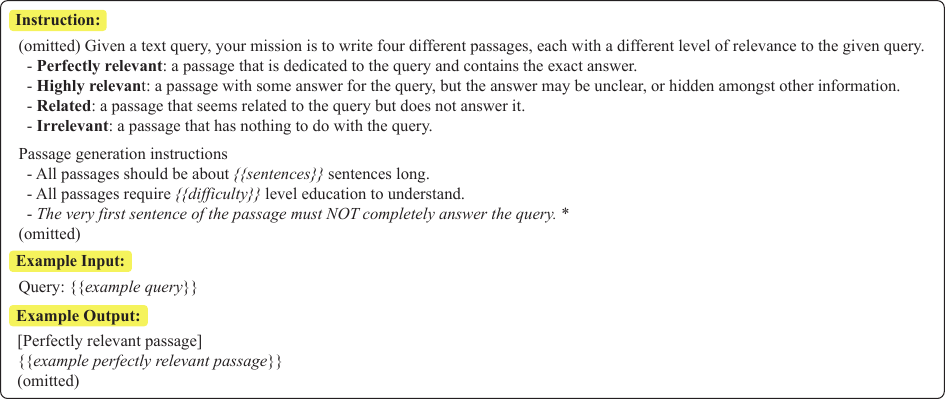}
  \caption{To create a multi-level ranking context for dense retrieval training, we prompt the LLM to sequentially generate four passages with graduated relevance levels for each query. To generate diverse passages, we randomly sample the value of \rinlinecode{\{\{sentences\}\}} and \rinlinecode{\{\{difficulty\}\}} for each prompt. To avoid easy-to-identify passages, we include the instruction with ``*'' in the prompt for a random subset of queries. See~\cref{sec:app_prompting} for details.}
  \label{fig:main_prompt}
\end{figure*}
\section{Related Work}
\label{sec:rel_work}

\paragraph{Dense Retrieval Training}
Retrieval training pipelines have improved significantly by addressing various limitations of IR datasets.
To better delineate the positive and negative regions, many have proposed using a large number of random in-batch negatives~\citep{karpukhin-etal-2020-dense} and similarly using existing retrievers to mine hard-to-detect negatives for each query~\citep{qu-2021-rocketqa,xiong2020approximate,zhan2021optimizing}.
Some have also used existing retrievers to filter out the unannotated positives during hard negative mining~\citep{moreira2024nv}.
However, the fundamental limitation remains: binary relevance labels provide a crude approximation of the ranking task.

\paragraph{Ranking Context}
The benefits of a multi-level ranking context are well established in the learning-to-rank (L2R) literature before the advent of transformer-based retrievers~\citep{cao2007learning,ai_learning_2019, ai_learning_2018}.
Most L2R works use 4 to 6 levels of relevance during training~\citep{qin_letor_2010, pmlr-v14-chapelle11a, dato_fast_2016} and hundreds of annotated documents per query, compared to current large-scale datasets, which only provide a binary definition of relevance, and mostly a single positive document.
As a result, the impact of multiple relevance levels for training large transformer models is largely unexplored, except for a few limited attempts.
For example, \citet{zerveas_coder_2022,zerveas-etal-2023-enhancing} use a large number of mined documents per query and label propagation based on a custom metric to show that even modern retrievers benefit from a rich ranking context.
However, their progress is fundamentally constrained by the limitations of available datasets.

\paragraph{Data Annotation}
Pseudo-labeling with cross-encoders is one approach for obtaining fine-grained relevance judgments~\citep{hashemi2023dense,wang2021gpl,zeng2022curriculum,faggioli2023perspectives,lee2024nv}.
However, because of the large corpus sizes, only a small set of retrieved documents is annotated for each query.
Since existing datasets contain many unannotated positives~\citep{qu-2021-rocketqa}, selecting a small set of candidate documents that covers a wide range of relevance levels is challenging, which reduces the annotation diversity for each query (\cref{sec:apndx_pseudo_lbl}).
Moreover, pseudo-labeling requires abundant data, which is not available for niche domains or novel applications like tool retrieval~\citep{qu2024towards}, or retrieval with reasoning and instructions~\citep{weller2025rank1,shao2025reasonir,su2024bright}.

Recent works have used LLMs for judging relevance in various setups~\citep{khramtsova2024leveraging,thomas2024large,faggioli2023perspectives,balog2025rankers,jin2024apeer,chen2024attention}.
However, the costs limit the scale of relevance judgments with LLMs.
Often, LLMs are used to only rerank the retrieved documents for a small number of test queries~\citep{zhuang2024setwise,qin2023large,sun2023chatgpt,ma2023zero}.
LLMs are also used to create small tests for evaluation or to judge the quality of other models~\citep{rahmani2025judging}.
Furthermore, a few works have used a small set of LLM annotations to fine-tune downstream rerankers~\citep{pradeep2023rankvicuna,pradeep2023rankzephyr}.
In addition to the extra costs, all the aforementioned problems for selecting candidate documents also apply to data annotation with LLMs.

Finally, it is possible to create fine-grained relevance data from search engine logs~\citep{rekabsaz_tripclick_2021}.
However, this also faces major challenges for rare queries or novel variants of the retrieval task where existing data is scarce (see \cref{sec:search_logs}).

\paragraph{Synthetic Data Generation}
IR pipelines have integrated synthetic data in different ways.
A popular approach is to create synthetic queries for existing passages~\citep{dai2022promptagator, bonifacio2022inpars, jeronymo2023inpars, alaofi2023can,lee2024geckoversatiletextembeddings}.
Another approach is to enhance the quality of existing queries~\citep{wang2023query2doc,shen2023large, jagerman2023query, rajapakse2023improving, Anand2023ContextAQ, li2024can, Dhole2024GenQREnsembleZL, Zhang2024ExploringTB}.
More recently, synthetic data has played an important role in training dense retrievers with reasoning~\citep{shao2025reasonir} and instruction-following~\citep{weller2024promptriever,asai2022task,wang-etal-2024-improving-text} capabilities.
Despite this diversity, all existing works generate synthetic data only with binary relevance levels, which inherits many problems of existing IR datasets discussed in~\cref{sec:intro}.
By contrast, we use the flexibility of LLMs to overcome the limitations of real data and generate multi-level ranking contexts, which are more suitable for training dense retrievers.

\section{Synthetic Ranking Context for List-wise Training (SyCL)}
\label{sec:method}

To better approximate the inference objective, we propose to train dense retrievers on passages with multiple levels of relevance, thus creating a rich \textit{multi-level ranking context} for each query.
Since most large-scale IR datasets only provide passages with binary ground truth labels, we use LLMs to generate passages with graduated relevance levels for each query (\cref{sec:data_gen}).
Additionally, we propose to use Wasserstein distance as a loss function that uses graded relevance labels and multiple positives per query more effectively than alternative list-wise losses (\cref{sec:training}).

\subsection{Multi-level Ranking Context}
\label{sec:data_gen}
We leverage open-source LLMs to generate multi-level ranking contexts for the MS~MARCO training queries at scale.
We use the official TREC Deep Learning\footnote{https://trec.nist.gov/data/deep2019.html} relevance guidelines to prompt LLMs to write passages that answer each query at four different levels of relevance: perfectly relevant, highly relevant, related, and irrelevant (\cref{fig:main_prompt}).
See \cref{sec:app_prompting} for the exact prompt.

For ranking, the \textit{relative} relevance of passages is the most important.
Even with clear instructions, when asked to generate passages of a specified relevance level without references, the LLM is not aware of how each will compare to other independently generated documents of the same, higher, or lower specified relevance to the same query.
Thus, we prompt the LLM to generate all four passages for each query \textit{sequentially} in the same inference session.
This allows the LLM to gradually decrease the relevance of each generated passage relative to already generated passages in its context in order to achieve the correct ranking order.
To help the LLM better understand the task, we provide one in-context example consisting of a query and four passages, i.e., one passage for each relevance level.

\paragraph{Corpus Diversity} Additionally, the in-context example reduces the distribution shift between the synthetic and real passages in terms of attributes like style.
Without examples, our synthetic documents tend to be distinctly clearer and more direct than real passages.
Hence, to increase the diversity of synthetic passages, for each prompting instance, we select a different in-context example from TREC DL 2023, which is meant for the new version of the MS~MARCO dataset (v2) and not used for training or evaluation by recent works.
Specifically, we randomly sample one of the 82 queries and one of its corresponding passages for each level as the in-context example.
This requires a very small number of ground truth labels (328 labeled passages in total), and compared to the scale of annotation in MS~MARCO (more than 500k annotated queries), the incurred cost is negligible.

Moreover, similar to~\citet{wang-etal-2024-improving-text}, for each prompt, we use templates to specify a randomly sampled passage length and difficulty level.
We also noticed that the LLM tends to provide the exact answer to the query in the very first sentence of the perfectly relevant passages, making them easily identifiable.
To prevent this, we explicitly instruct the LLM to avoid this in a random subset of prompts.
See Appendix~\ref{sec:app_prompting} for more details.

We use simple text processing to extract the four passages from the LLM response and assign them sequential labels, \texttt{\{3,2,1,0\}}, based on the specified relevance level for which they were generated.

\subsection{Training with Multiple Levels of Relevance}
\label{sec:training}

To effectively leverage the generated multi-level ranking contexts, we propose using the 2-Wasserstein distance as loss function.
Although it has been used in retrieval pipelines as a distance in different roles, e.g., regularization~\citep{yu-etal-2020-wasserstein}, to the best of our knowledge we are the first to propose it as a relevance loss function for training dense retrievers.
We use a differentiable analytical expression of Wasserstein distance that can be efficiently computed when comparing two Gaussian distributions~\citep{mathiasen2020backpropagating}.
Although neither our ground truth nor estimated score distributions are Gaussian, this approximation outperforms the most popular list-wise loss functions.
For two multivariate Gaussian distributed inputs $X \sim \mathcal{N}(\mu_x, C_x)$ and $Y \sim \mathcal{N}(\mu_y, C_y)$, where $\mu$ and $C$ are the mean and covariance of each distribution, we calculate the 2-Wasserstein distance as follows:
\begin{equation*}
   \operatorname{D}(X, Y) = \|\mu_x - \mu_y\|^2 - \operatorname{tr}(C_x + C_y - 2(C_xC_y)^{\frac{1}{2}}) \, .
\end{equation*}
During training, we present labels and predicted scores as matrices $H$ and $\hat{H}$ of shape \texttt{(batch size, ranking context size)} and minimize $\operatorname{D}(H, \hat{H})$.

Compared to KL divergence, which has been used as a multi-level list-wise loss function~\citep{zerveas-etal-2023-enhancing,wang2022text}, the Wasserstein distance has the following main advantages.
First, distributing probability mass over candidate documents is penalized according to their ground-truth score distance from the ground-truth target document: assigning some probability mass to a document with g.t. label 0 instead of the correct document with g.t. 3 is penalized more strongly than assigning it to a document with g.t. label 2.
By contrast, the KL divergence is insensitive to this relative distance in the estimated score distribution.
As long as the g.t. relevant document (or any other document) is not assigned its due g.t. probability mass in the estimated score distribution, it will be penalized the same regardless of where this probability mass goes.
Second, it is computed by comparing the ground truth and estimated score distributions across documents of the entire batch, not only across those in the context of a single query.
We hypothesize that this acts as a regularization, e.g., granting resilience to the range of score values or outliers.

\begin{table*}[t]
\centering
\setlength{\tabcolsep}{4pt}
\resizebox{\textwidth}{!}{
\begin{tabular}{@{}lccc|cccccc@{}}
\toprule
\textbf{nDCG@10} & \textbf{DL19} & \textbf{DL20} & MM Dev & FEVER & HotpotQA & FiQA & NQ & Quora & Touche   \\
\cmidrule(lr){2-10}
Base Contriever (BC)                      & 45.5\hab{} & 44.8\hab{} & 20.6\hab{} & 66.8\hab{} & 48.2\hab{} & 24.6\hab{} & 25.4\hab{} & 83.5\hab{} & 18.6\hab{}  \\ \midrule
BC + \infoncesynth{}                      & 55.3\hab{} & 51.5\hab{} & 26.3\hab{} & 68.0\hab{} & 46.4\hab{} & 26.8\hab{} & 33.2\hab{} & 75.8\hab{} & 15.0\hab{}     \\

\textcolor{purple}{BC + \wssynth{}}       & 59.6\vab{} & 59.8\vab{} & 30.2\vab{} & 81.8\vab{} & 57.2\vab{} & 27.3\vaa{} & 41.9\vab{} & 83.3\vbb{} & 20.3\vbb{}     \\ \cdashlinelr{1-10}

BC + \infoncereal{}                       & 63.0\hab{} & 61.2\hab{} & 34.2\hab{} & 69.6\hab{} & 59.8\hab{} & 29.1\hab{} & 42.8\hab{} & 81.7\hab{} & 14.6\hab{}     \\
\textcolor{purple}{BC + \wswithreal{}}    & 63.2\hab{} & 61.6\hab{} & 32.9\hab{} & 80.6\hab{} & 59.8\hab{} & 30.0\hab{} & 42.5\hab{} & 83.7\hab{} & 16.8\hab{}     \\
\end{tabular}
}
\resizebox{\textwidth}{!}{
\begin{tabular}{@{}lccccccccc@{}}
\toprule
\textbf{nDCG@10} & \makecell{CQADup\\Android} & Scidocs & \makecell{Climate\\FEVER} & DBPedia & \makecell{TREC\\COVID} & Scifact & NFCorpus & ArguAna & \makecell{\textbf{BEIR}\\\textbf{Avg}} \\ \cmidrule(l){2-10} 
Base Contriever (BC)                       & 37.5\hab{} & 15.1\hab{} & 15.2\hab{} & 29.4\hab{} & 27.7\hab{} & 63.9\hab{} & 32.4\hab{} & 31.4\hab{} & 37.1\hab{}    \\ \midrule
BC + \infoncesynth{}                       & 35.0\hab{} & 15.1\hab{} & 21.4\hab{} & 32.0\hab{} & 26.6\hab{} & 62.5\hab{} & 31.5\hab{} & 26.4\hab{} & 36.8\hab{}    \\

\textcolor{purple}{BC + \wssynth{}}        & 39.0\vbb{} & 16.4\vab{} & 27.0\vab{} & 36.7\vab{} & 52.7\vab{} & 62.0\hab{} & 31.8\hab{} & 28.2\vab{} & 43.2\hab{}    \\ \cdashlinelr{1-10}

BC + \infoncereal{}                        & 38.2\hab{} & 16.2\hab{} & 18.3\hab{} & 37.6\hab{} & 34.0\hab{} & 65.1\hab{} & 31.5\hab{} & 33.6\hab{} & 40.9\hab{}    \\
\textcolor{purple}{BC + \wswithreal{}}     & 40.5\hab{} & 16.0\hab{} & 25.5\hab{} & 38.9\hab{} & 51.2\hab{} & 66.6\hab{} & 33.0\hab{} & 33.0\hab{} & 44.2\hab{}    \\ \bottomrule
\end{tabular}
}
\caption{Ranking effectiveness (nDCG@10). Base Contriever (BC): self-supervised Contriever model. `BC +' denotes the fine-tuning setting in terms of \textbf{loss function}: InfoNCE / Wasserstein (WS), and \textbf{type of data}: real data from the MS~MARCO training set with annotated positives and mined hard negatives (Real) / fully synthetic multi-level documents (Synth.) / combination. DL19, DL20, and MM Dev are the TREC DL 2019, TREC DL 2020, and Dev evaluation sets of MS~MARCO. Evaluation on the rest of sets is zero-shot. Symbols $^a$ and $^b$ denote a statistically significant difference (paired $t$-test) with $p < 0.05$ when compared to BC and BC + \infoncereal{}, respectively. \textcolor{purple}{Purple: SyCL, our method.}}
\label{tab:main}
\end{table*}

\section{Experiments}
\label{sec:exp}

Our experiments demonstrate the effectiveness of synthetic multi-level ranking contexts and the Wasserstein loss for training dense retrievers.
First, without using any real documents or annotations, SyCL fine-tuning improves the performance of self-supervised dense retrievers.
Second, we show that the Wasserstein loss with multiple levels of relevance outperforms InfoNCE using the same queries and passages.
Third, we find that SyCL training only on synthetic documents performs similarly to contrastive training with real data of the same size on TREC DL, while on average, it outperforms it in terms of out-of-domain generalization on BEIR.
Overall, we achieve the best ranking effectiveness when incorporating existing real data into our synthetic multi-level ranking context.
Through additional analytical experiments, we show the individual impact of the Wasserstein loss and graduated relevance labels.
Finally, we inspect different components of our data generation pipeline and find that even smaller, 32B-scale LLMs can generate high-quality data comparable to larger 70B-parameter models.

\subsection{Setup}
\label{sec:setup}

\paragraph{Training}
We use Llama 3.3 70B~\citep{dubey2024llama} to generate one passage for each level of relevance (i.e., ranking context size of four) for training queries of the MS~MARCO dataset (total of \textasciitilde 2M passages).
During training, we use all passages corresponding to other queries in the batch as level zero passages in the multi-level ranking context of a given query.
We use the unsupervised Contriever~\cite{izacard2021contriever} model as our base model.
See~\cref{sec:other_models} for experiments with other models.
More implementation details are provided in~\cref{sec:imp_details}.

\paragraph{Evaluation}
For in-domain evaluations, we use the TREC DL 2019, TREC DL 2020, and Dev set of the MS~MARCO dataset.
To evaluate how well our model performs in the real world, we use the 14 publicly available datasets in the BEIR benchmark~\citep{thakur2021beir} for out-of-domain evaluation.
To simplify our BEIR evaluations for duplicate question retrieval, we only use the Android subforum of the CQADupStack dataset.

\subsection{Results}

\cref{tab:main} shows our main results on the effectiveness of using a synthetic multi-level ranking context with the Wasserstein loss to train dense retrievers.

SyCL significantly improves the performance of the unsupervised Contriever model for both in-domain evaluation on the MS~MARCO dataset and out-of-domain evaluation on the BEIR benchmark.
In terms of nDCG@10, our method improves the base model performance by 6.2 across BEIR, 14.1 on TREC DL19, and 14.9 on TREC DL20.

Notably, for in-domain evaluation, the performance boost for the DL19 and DL20 sets is more significant than that of the Dev set (9.7).
This is expected: MS~MARCO Dev is extremely sparsely annotated (mostly, one positive per query) and missing most real positive documents.
Compared to contrastive training with a single positive, a training method like ours teaches the model to distribute relevance scores among more documents in the ranking context (see~\cref{fig:pred_scores} in the Appendix).
Consequently, it has a much higher probability of assigning a high score to documents other than the annotated positive, and the chance for the latter to be displaced to lower ranks increases.
Therefore, the question is whether the documents displacing the ground-truth positive are indeed relevant. Qualitative inspection of ranked documents (\cref{tab:retrieval_samples} in the Appendix) and evaluation on more densely annotated sets (\cref{tab:main}) indicate that the answer is affirmative and may explain the difference in performance improvements.
DL19 and DL20 additionally provide multi-level relevance labels, which helps to better evaluate the fine-grained ranking capabilities of retrievers.

\textbf{Multi-level ranking context with Wasserstein loss uses the same data more effectively than InfoNCE.}
For an apples-to-apples comparison with the standard contrastive training, we train the model with InfoNCE loss using the same synthetic passages (\infoncesynth{} in \cref{tab:main}).
For this, we use the passages from levels 3 and 2 as positives and passages from levels 1 and 0 as negatives.
Although both setups use the same queries and passages, multi-level ranking context with Wasserstein loss uses the data more effectively and clearly outperforms contrastive training.

\begin{table}[t]
\centering
\resizebox{\columnwidth}{!}{
\begin{tabular}{@{}lcccc@{}}
\toprule
            & DL19  & DL20  & MS Dev & BEIR  \\ \cmidrule(l){2-5} 
Synth. Binary + WS     & 48.9 & 48.7 & 22.1  & 40.5 \\
Synth. Multi-Level + WS & 59.6 & 59.8 & 30.2  & 43.2 \\ \midrule
Real Binary + WS      & 50.6 & 47.4 & 23.6  & 36.6 \\
Real Binary + InfoNCE & 63.0 & 61.2 & 34.2  & 40.9 \\ 

\bottomrule
\end{tabular}
}
\caption{Top: nDCG@10 of models trained with Wasserstein loss on the same synthetic data with binary ($\{1,0\}$) and graduated ($\{3,2,1,0\}$) relevance labels. Bottom: nDCG@10 of models trained with Wasserstein and InfoNCE loss on real data with binary labels.}
\label{tab:bin_vs_multi}
\end{table}

To evaluate contrastive training with real data, we use the human-annotated positives and two hard negatives mined by BM25 to match the number of negatives in synthetic data (\infoncereal{} in \cref{tab:main}).
Although training with real, labeled documents leads to slightly better performance for in-domain evaluation on MS~MARCO, training exclusively on synthetic documents performs comparably.
On the other hand, SyCL better generalizes to out-of-domain datasets in the BEIR benchmark and outperforms real data by 2.3 nDCG@10 on average.
This indicates better robustness to distribution shift and unseen data, which has been argued to be the most important attribute of IR methods for real-world applications~\citep{thakur2021beir}.

\textbf{Augmenting real data with multi-level synthetic passages further improves performance.}
To benefit from both real and synthetic data, we assign relevance levels 3 and 1 to positive and negative real passages respectively, and incorporate them into the synthetic multi-level ranking context.
Combining synthetic and real data improves SyCL's ranking effectiveness on DL19 from 59.6 to 63.2, and on DL20 from 59.8 to 61.4.
Compared to training with real data and the InfoNCE loss, training with SyCL on the combined data improves nDCG@10 scores from 40.9 to 44.2 on the BEIR benchmark. 
Adding real data seems to slightly degrade performance on MS~MARCO Dev, which we attribute to its extremely sparse annotation (see our earlier discussion in this section).

\section{Additional Analysis}
\label{sec:additional_analysis}

\textbf{Fine-grained relevance levels are necessary for achieving good performance.}
To separate the impact of using multiple relevance levels from the Wasserstein loss, we repeat our main experiment with the Wasserstein loss but instead use binary labels.
We assign relevance levels 1 and 0 to more relevant (levels 3 and 2) and less relevant (levels 1 and 0) synthetic passages, respectively (\cref{tab:bin_vs_multi} top).
Even with the same data and loss function, fine-grained relevance levels are necessary for achieving good performance: using binary relevance levels instead decreases the boost in performance by 4.0 nDCG@10 on average across all sets.\\
Although our main comparison is between binary and multi-level synthetic data, we also experiment with fine-tuning on real binary data using Wasserstein loss (\cref{tab:bin_vs_multi} bottom).
For real data with binary labels, InfoNCE performs better than Wasserstein loss.
Therefore, without a multi-level ranking context, the Wasserstein loss by itself does not explain the performance gains of our approach, which reinforces our main claim: the combination of synthetic multi-level data and the Wasserstein loss is particularly effective for fine-tuning dense retrievers.

\begin{table}[t]
\centering
\resizebox{\columnwidth}{!}{
\begin{tabular}{@{}lcccc@{}}
\toprule
Loss       & DL19  & DL20  & MS Dev & BEIR \\ \midrule
Approx.~nDCG & 54.7 & 52.5 & 27.8    & 39.1    \\
RankNet    & 54.3 & 48.9 & 24.6    & 35.3    \\
ListNet    & 56.9 & 55.4 & 27.5     & 42.2     \\
KL-div         & 56.1 & 54.9 & 27.4    & 42.1    \\ \midrule
Wasserstein         & 59.6 & 59.8 & 30.2    & 43.2    \\ \bottomrule
\end{tabular}
}
\caption{Performance (nDCG@10) of models trained on multi-level synthetic data with different list-wise losses.}
\label{tab:other_losses}
\end{table}

\textbf{Wasserstein loss is more effective than other list-wise loss functions.}
We compare our proposed Wasserstein loss against other list-wise loss functions that can take advantage of multiple levels of relevance (\cref{tab:other_losses}).
We evaluate the Approximate NDCG (a smooth, differentiable approximation of the nDCG metric)~\citep{qin2010general}, RankNet~\citep{burges2005learning}, and ListNet~\citep{cao2007learning} loss functions, which have been used extensively in learning-to-rank approaches before the advent of dense retrieval.
We also evaluate the KL divergence, which is often used for model distillation
but has also been used for training with a multi-level ranking context~\citep{zerveas_coder_2022, zerveas-etal-2023-enhancing}.
Except for RankNet, all other loss functions take advantage of multiple levels of relevance and outperform the binary InfoNCE loss.
However, the Wasserstein loss is the most effective and provides significant gains over the next best loss function (ListNet).

\begin{table}[t]
\centering
\resizebox{\columnwidth}{!}{
\begin{tabular}{@{}lcccc@{}}
\toprule
            & DL19  & DL20  & MS Dev & BEIR  \\ \cmidrule(l){2-5} 
Direct Synth. Binary      & 47.8 & 40.4 & 20.5  & 36.6 \\
Approximated Synth. Binary & 58.4 & 52.0 & 	26.0  & 37.6 \\

\bottomrule
\end{tabular}
}
\caption{Performance using InfoNCE loss with binary passages directly generated by the LLM and approximated binary passages (i.e., multi-level passages with with binary labels)}
\label{tab:direct_synth_binary}
\end{table}

\textbf{Generating multi-level synthetic data is better even for binary training.}
In our experiments thus far, we approximate binary synthetic data by using the same multi-level synthetic passages but converting the labels from multi-level to binary.
This helps us study the impact of label granularity without confounding factors like variation in passage content.
We now evaluate directly generating binary data using Qwen 2.5 32B and report the results (nDCG@10) in \cref{tab:direct_synth_binary} (exact prompt in \cref{sec:app_prompting}).
The bespoke binary synthetic passages perform even worse than the simulated binary passages used in our main experiments.
This further strengthens our claim about the merits of generating multi-level synthetic data.
We hypothesize that when prompted to generate passages with multiple levels of relevance, the LLM more precisely controls the content of each passage in order to meet the relevance requirements, resulting in more nuanced and challenging passages.

\textbf{Even with a strict interpretation of relevance, models trained with SyCL outperform BM25 without using any real passages.}
We show that even under a strict interpretation of relevance labels, our method outperforms BM25 without using any real passages or their annotations (\cref{tab:without_one} in the Appendix).
Following TREC guidelines, we exclude passages with relevance label 1 for the strict evaluation setup.
To be a viable approach for practical applications, dense retrievers should at least perform better than BM25, which does not require any training and still achieves strong performance.
However, most dense retrieval methods fail to outperform BM25 without additional fine-tuning on labeled training data.
Recently, \citet{wang2022text} managed to outperform BM25 without using any labeled data.
However, they resorted to a complex multi-stage training pipeline to achieve this, while we use synthetic data to better capture the ranking objective during training with a simple pipeline.

\begin{figure}[t]
  \centering
  \includegraphics[width=\columnwidth]{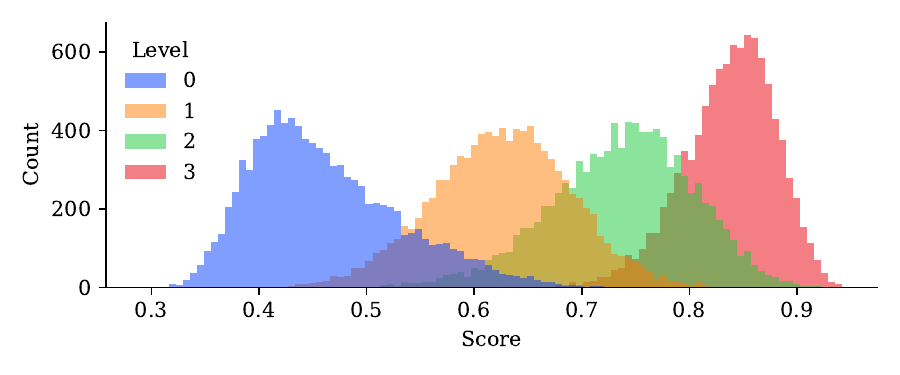}
  \caption{Distribution of the similarity scores between queries and the corresponding synthetic passages for each relevance level.}
  \label{fig:scores_per_level}
\end{figure}
\begin{table}[t]
\centering
\resizebox{\columnwidth}{!}{
\begin{tabular}{@{}lcccc@{}}
\toprule
              & DL19  & DL20  & MS Dev & BEIR  \\ \cmidrule(l){2-5} 
Llama 3.3 70B & 59.6 & 59.8 & 30.2  & 43.2 \\
Qwen 2.5 72B  & 60.0 & 57.9 & 30.3  & 42.4 \\
Qwen 2.5 32B  & 61.1 & 56.9  & 29.3  & 42.9 \\ \bottomrule
\end{tabular}
}
\caption{Performance of models trained with synthetic data generated by different LLMs.}
\label{tab:other_llms}
\end{table}

\begin{table*}[t]
\small
\centering
\setlength{\tabcolsep}{4pt}
\resizebox{\textwidth}{!}{
\begin{tabular}{p{0.06\textwidth}p{0.93\textwidth}}
\toprule

Query
&
\tt
\scriptsize
what do cells do?
\\
\midrule
Level 3

&
\tt
\scriptsize
Cells are the basic structural and functional units of living organisms. They are responsible for carrying out various functions necessary for life, including growth, reproduction, metabolism, and response to stimuli. Cells take in nutrients, convert them into energy, and eliminate waste products. They also contain genetic material, known as DNA, which carries information necessary for the cell's development and function. Additionally, cells communicate with each other to coordinate their activities and maintain overall health.
\\
\midrule
Level 2
&
\tt
\scriptsize
The human body is made up of trillions of cells, each with its own unique characteristics and functions. Some cells, such as muscle cells, are specialized for movement and contraction, while others, like nerve cells, are responsible for transmitting signals and controlling various bodily functions. Cells also have the ability to adapt to their environment, change shape, and even migrate to different parts of the body. For example, immune cells can recognize and respond to foreign substances, helping to protect the body from infection and disease.
\\
\midrule
Level 1
&
\tt
\scriptsize
Tissues are groups of similar cells that work together to perform specific tasks. Organs are structures made up of two or more types of tissues that function together to maintain the overall health of an organism. The study of the structure, behavior, and function of tissues and organs is essential for understanding how the body works and how diseases develop. By examining the relationships between different tissues and organs, researchers can gain insights into the complex processes that occur within living organisms.
\\
\midrule
Level 0
&
\tt
\scriptsize
The city of Paris is known for its stunning architecture, rich history, and vibrant cultural scene. Visitors from around the world flock to the City of Light to marvel at iconic landmarks like the Eiffel Tower, Notre-Dame Cathedral, and the Louvre Museum. The city's charming streets and alleys are lined with quaint cafes, boutiques, and restaurants, offering a unique and unforgettable experience for travelers. Whether you're interested in art, fashion, or food, Paris has something to offer everyone.
\\

\bottomrule
\end{tabular}
}
\caption{Synthetic multi-level ranking context generated for one query. The LLM successfully follows the defined relevance levels and generates passages with correct relative similarity to the given query.}
\label{tab:sample_data}
\end{table*}

\subsection{Analysis of Synthetic Data Generation Process}

\textbf{LLMs successfully follow the definition of relevance levels.}
To check if synthetic passages adhere to their corresponding relevance level, we use a high-quality embedding model, \texttt{e5-mistral-instruct}~\citep{wang2023improving}, to measure the similarity between 10,000 randomly selected queries and their corresponding synthetic documents.
\Cref{fig:scores_per_level} reports the distribution of similarity scores for passages in each relevance level.
We find that the generator LLM understands the relevance levels and appropriately decreases the relevance between the query and generated document based on its pre-specified target level and the documents sequentially generated before it.

We provide a sample of the generated passages in~\cref{tab:sample_data},
which shows that the LLM first generates a positive passage that fully answers the query and then, with some nuanced changes, creates a less relevant positive passage that provides a partial answer to the query, and similarly keeps reducing the relevant content for the other two less relevant passages in the context.

\textbf{Small LLMs also generate high-quality data.}
To understand the impact of the LLM on the quality of the synthetic data, we also generate data with two other LLMs, Qwen 2.5 72B and Qwen 2.5 32B~\citep{qwen25}, and use it to train the retriever similar to our main experiments (\cref{tab:other_llms}).
For in-domain evaluation, data generated with larger LLMs leads to better performance on DL20 and Dev splits of MS~MARCO.
However, for out-of-domain evaluation on BEIR datasets, data generated with the smaller Qwen 2.5 32B leads to performance similar to data generated with Llama 3.3 70B.
Although recent works use 70B scale public or larger proprietary models~\citep{wang2023improving,weller2024promptriever}, our results show that data generated with larger models does not always lead to better performance.
See~\cref{sec:data_size} for experiments using combined data from all LLMs.

\begin{table}[t]
\centering
\setlength{\tabcolsep}{4pt}
\resizebox{\columnwidth}{!}{
\begin{tabular}{@{}lcccc@{}}
\toprule
                 & DL19  & DL20  & MS Dev & \multicolumn{1}{l}{BEIR} \\ \cmidrule(l){2-5} 
Full             & 59.6 & 59.8 & 30.2  & 43.2                    \\
No in-context example    & 59.5 & 60.0 & 29.9  & 42.8                    \\
No random variation & 59.9 & 58.3 & 29.8  & 42.8                    \\ \bottomrule
\end{tabular}
}
\caption {Impact of prompt design on retrieval performance. Full: our main prompt. No IC example: prompt without in-context examples. No random variation: prompt without randomly sampled instructions (e.g. length requirement).}
\label{tab:diversity_ablation}
\vspace{-3mm}
\end{table}

We investigate the impact of the in-context example and the randomly selected instructions (e.g., length) on the quality of the synthetic data.
We create two alternative prompts, one without the in-context examples and the other without the randomly selected instructions, and use the resulting data for training (\cref{tab:diversity_ablation}).
Although both techniques contribute to the quality of synthetic data, in-context examples are more important, especially for out-of-domain generalization to BEIR datasets.
\section{Conclusion}

In this work, we introduce SyCL, a novel method that first uses LLMs to generate rich multi-level ranking contexts and then the Wasserstein distance to train retrievers with multiple levels of relevance.
We show that LLMs can successfully generate synthetic data with graduated relevance levels, significantly improving the effectiveness of unsupervised retrievers.
When using the same synthetic queries and passages, SyCL utilizes the available data more effectively and performs better than training with binary relevance labels.
SyCL can also combine real and synthetic datasets to further improve performance.
Moreover, we show that Wasserstein distance is more effective at fine-tuning transformer-based retrievers with graduated relevance labels and performs better than the usual list-wise loss functions.
Our results show that generating multiple passages with graduated relevance levels is a better approach to synthetic data generation for IR than generating the standard positive and negative passages.
These results encourage future work to explore synthetic data generation methods that are better suited for information retrieval tasks, going beyond the binary definition of relevance.
\section*{Limitations}

\paragraph{LLM Capabilities}
Similar to other works on synthetic data generation, our work is limited by the capabilities of LLMs.
For instance, data generation for specialized domains could pose a challenge for existing LLMs, especially at smaller scales.
Considering the progress in generating synthetic instruction tuning data for specialized domains~\citep{nayak2024learning}, we encourage future work to explore opportunities to expand applications of synthetic ranking data to specialized domains as well.

\paragraph{Dependency on Existing Queries}
Our work requires the availability of a collection of user queries in the target domain.
For many domains, a large collection of user queries is already provided by existing datasets or can be collected from online forums like Reddit or from users' conversation history with LLM assistants.
However, for very rare applications where none of these resources is available, we encourage future work to explore the combination of our work with synthetic query generation techniques~\citep{wang-etal-2024-improving-text}.
However, generating a large collection of queries from scratch also comes with its own challenges.
While there are many frequently occurring queries, 70\% of (distinct) queries occur only once~\citep{brenes_stratified_2009}.
Therefore, the LLM would be challenged to imagine representative user queries in most situations.

\paragraph{More Fine-grained Relevance Levels}
Moreover, we assume that LLMs understand the difference between relevance levels and can generate suitable data accordingly.
We show experimentally that this is, in fact, the case, and LLMs successfully generate documents with four different relevance levels.
However, we speculate that if we increase the number of relevance levels, after a certain point, the differences would be too nuanced for existing LLMs to recognize and follow.
We encourage future work to study the limitations of existing LLMs in terms of understanding nuanced semantic differences through instructions and also explore more advanced approaches for controlling the semantic similarity of the generated documents.

\section*{Ethical Considerations}

Since we use the MS MARCO training queries to guide the data generation process, our synthetic data might inherit the social biases and ethical concerns related to the MS MARCO dataset.
Moreover, similar to other works on synthetic data generation, our data also inherits the social biases and ethical concerns related to the LLM used for generating the synthetic documents.
Although we did not observe any harmful content during the course of this project, a principled analysis of social biases, factual correctness, and other ethical concerns is needed before use in sensitive real-world applications.

\section*{Acknowledgements}

This material is based upon work supported by the National Science Foundation under Grant No. RISE-2425380. Any opinions, findings, and conclusions or recommendations expressed in this material are those of the author(s) and do not necessarily reflect the views of the National Science Foundation. Disclosure: Stephen Bach is an advisor to Snorkel AI, a company that provides software and services for data-centric artificial intelligence.

\bibliography{misc/refs}

\appendix

\begin{figure}[h]
  \centering
  \includegraphics[width=\columnwidth]{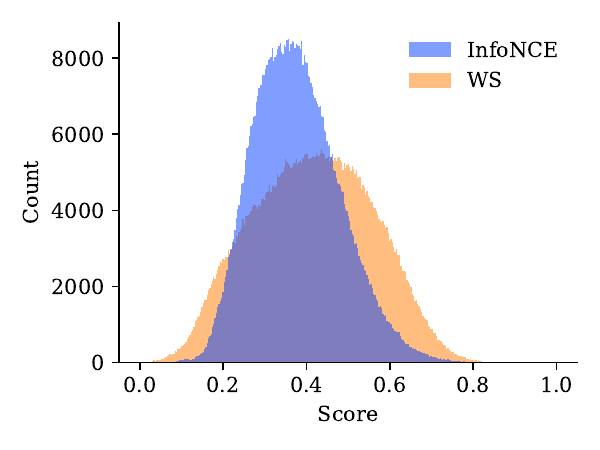}
  \caption{Distribution of the top 100 similarity scores across all Dev queries of MS~MARCO dataset by models trained with Wasserstein and InfoNCE losses. The model trained with multiple relevance levels learns a more fine-grained notion of relevance.}
  \label{fig:pred_scores}
\end{figure}

\begin{table}[h]
\centering
\begin{tabular}{@{}lcc@{}}
\toprule
\textbf{nDCG@10}                     & DL19  & DL20  \\ \midrule
BM25~\cite{yang_anserini_2017}  & 41.7 & 41.2\\
Base Contriever (BC) & 37.6 & 36.9 \\ \midrule
BC + \infoncesynth{}       & 48.0 & 45.0 \\
\textcolor{purple}{BC + \wssynth{}}                   & 52.6 & 53.4 \\ \midrule
BC + \infoncereal{}         & 57.7 & 55.4 \\
\textcolor{purple}{BC + \wswithreal{}}   & 56.6 & 54.6 \\ \bottomrule
\end{tabular}
\caption{Evaluation excluding passages with label 1 (Related), as per the official TREC guidelines}
\label{tab:without_one}
\end{table}

\section{Discussion}
\label{sec:discussion}

\subsection{Search Engine Logs}
\label{sec:search_logs}

Although it is feasible to create IR datasets with graduated relevance labels using search engine click logs~\citep{rekabsaz_tripclick_2021}, it comes with significant technical and practical challenges.
Technically, extensive search engine logs are only available for popular domains, leaving out niche applications (e.g., climate research).
Even for popular domains, given the nature of click-through models, graduated relevance labels are only possible for frequent queries, and rare queries are left with sparse binary annotations~\citep{rekabsaz_tripclick_2021}.
Practically, search engine logs are valuable business assets and are only selectively released by large companies, which limits the coverage and quality of the resulting datasets.
By contrast, our method uses open-weight LLMs to generate large datasets with graduated relevance labels, which are applicable to many diverse domains and queries while being publicly accessible.

\subsection{Pseudo-labeling}
\label{sec:apndx_pseudo_lbl}

As discussed in \cref{sec:rel_work}, pseudo-labeling requires access to large amounts of existing data, which is not available for rare domains, new applications, and new variants of the retrieval task.
Moreover, pseudo-labeling depends on existing retrievers and cross-encoder, which limits the quality of the resulting data.
Besides, there is no existing retriever or cross-encoder with acceptable performance for recent tasks like tool retrieval~\citep{qu2024towards} or retrieval with reasoning~\citep{shao2025reasonir}.
Even beyond these issues, selecting a small collection of candidate documents for pseudo-labeling that covers a wide range of relevance levels for each query is difficult.
Simple methods like BM25 often fail to retrieve multiple relevant documents with nuanced differences.
On the other end of the spectrum, because existing datasets are sparsely annotated, good dense retrievers often select the unannotated positives as labeling candidates and do not capture slightly less relevant but still informative documents.

\begin{figure*}[h]
\centering

\begin{subfigure}{\textwidth}
\centering
\begin{subfigure}{77mm}
\centering
\includegraphics[width=\textwidth]{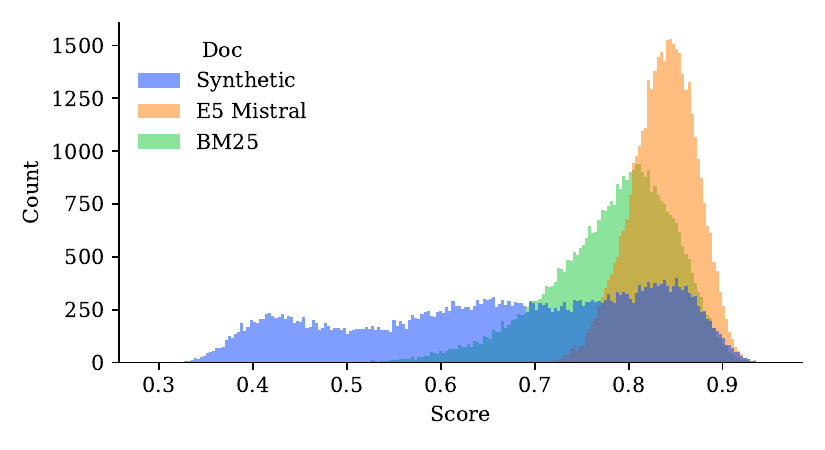}
\caption{}
\label{fig:pseudo_lbl_top4}
\end{subfigure}
\hspace{4mm}
\begin{subfigure}{77mm}
\centering
\includegraphics[width=\textwidth]{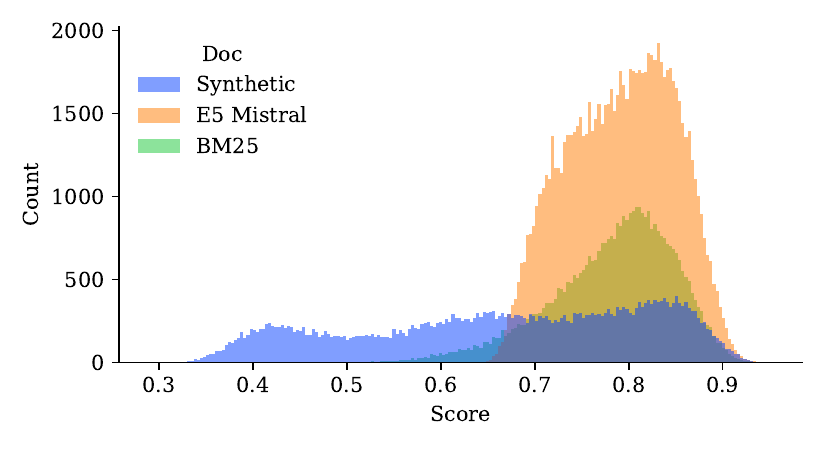} 
\caption{}
\label{fig:pseudo_lbl_top4plus}
\end{subfigure}
\end{subfigure}

\caption{Comparison of the distribution of similarity scores for synthetic documents and candidate documents selected for pseudo-labeling by BM25 and E5-Mistral-Instruct. a) For E5-Mistral-Instruct, we only select the the top-4 mined documents. b) For E5-Mistral-Instruct, we select both the top-4 and the the 90$^{th}$ to 95$^{th}$ mined documents for each query.}
\label{fig:pseudo_lbl_dists}
\end{figure*}

We randomly select 10,000 queries and measure the similarity of synthetic documents and pseudo-labeling candidates using \texttt{e5-mistral-instruct} (\cref{fig:pseudo_lbl_top4}).
We observe that synthetic documents cover a wide range of relevance levels, from irrelevant to perfectly relevant.
However, documents selected by BM25 are within a much narrower range of relevance levels. Most of them are relevant to the given query but not perfectly relevant.
Candidate documents selected by \texttt{e5-mistral-instruct} are within an even narrower range of relevance levels and are mostly unannotated positives.
Although with \texttt{e5-mistral-instruct}, we can ignore the top-ranked documents and choose less relevant documents, it does not significantly improve the diversity of annotations.
In \cref{fig:pseudo_lbl_top4plus}, we use \texttt{e5-mistral-instruct} and choose the top 4 documents (equal to the number of synthetic documents) as well as the 90$^{th}$ to 95$^{th}$ documents.
It definitely widens the similarity range of candidate documents, but it is still much more limited than synthetic documents.

In \cref{tab:pl_samples3}, we show the synthetic documents generated for a sample query as well as the candidate documents selected for pseudo-labeling by BM25 and \texttt{e5-mistral-instruct}.
For the synthetic documents, the LLM first answers the query directly in the most relevant document and then makes nuanced changes to decrease the relevancy of the answer for each subsequent document.
Finally, it generates a totally irrelevant passage for the last relevance level.
On the other hand, \texttt{e5-mistral-instruct} selects the unannotated positives for pseudo-labeling, which reduces the diversity of annotations (i.e., all candidates will be labeled as ``perfectly relevant'').
Finally, the documents selected by BM25 do not answer the query at all and are not useful for learning the nuanced differences between multiple relevant documents.

\subsection{Decoder-only Retrievers}

Although retrievers based on large LLMs, such as E5-Mistral-Instruct~\citep{wang2023improving} and RepLlama~\citep{ma2024fine}, have achieved significant performance improvements in academic setups, smaller BERT-sized models are still of extreme importance.
Inference costs are a major concern for information retrieval.
And, even the authors of E5-Mistral-Instruct emphasize the importance of smaller models.
Quote from~\citet{wang2023improving}: ''\textit{In comparison to the mainstream BERT-style encoders, the employment of LLMs, such as Mistral7B, for text embeddings results in a significantly increased inference cost.}`` and ''\textit{With regards to storage cost, our model is comparatively more expensive, with embeddings of 4096 dimensions.}``
Many practical applications involve millions, if not billions, of documents. Smaller BERT-sized retrievers are preferred in such cases. Even in academic setups, many recent RAG methods use BERT-based retrievers in their pipeline~\citep{gao2023retrieval}, which further emphasizes the importance of smaller models for dense retrieval.

Finally, although we do not use decoder-only retrievers in our work due to practical constraints, we expect that such retrievers would benefit even to a greater extent from our training methodology, as they would be more sensitive to the nuanced training signal offered by our multi-level ranking contexts.

\begin{table}[t]
\centering
\resizebox{\columnwidth}{!}{
\begin{tabular}{@{}lcccc@{}}
\toprule
              & DL19  & DL20  & MS Dev & BEIR  \\ \cmidrule(l){2-5} 
Llama 3.3 70B & 59.6 & 59.8 & 30.2  & 43.2 \\
All LLMs & 60.3 & 58.3 & 30.0 & 43.1 \\ \bottomrule
\end{tabular}
}
\caption{Performance of the model trained with synthetic data generated only by Llama 3.3 70B compared to the model trained on the combination of synthetic data generated by Llama 3.3 70B, Qwen 2.5 72B, and Qwen 2.5 32B.}
\label{tab:mix3}
\end{table}

\section{Impact of Data Size}
\label{sec:data_size}

We run additional experiments to investigate how increasing the number of synthetic passages for each query impacts the performance.
We combine the data generated by all three LLMs (i.e., Llama 3.3 70B, Qwen 2.5 72B, and Qwen 2.5 32B) and use it to train the base retriever, similar to our main experiments (\cref{tab:mix3}).
We find that increasing the size of the data does not have a noticeable impact on performance, which suggests that the quality of the data is more important than its quantity.
However, these results should be interpreted with caution since there could be other confounding factors, such as the calibration of ground-truth label values between the three different LLMs.
For instance, even if the level 3 document generated by one LLM is less relevant than the level 3 document generated by another LLM for the same query, we use label 3 for both of them in this experiment.
Making reliable conclusions about the impact of data size requires extensive experiments that control for this and other confounding factors.
We leave such analysis to future work. 

\begin{table}[h]
\centering
\resizebox{\columnwidth}{!}{
\begin{tabular}{@{}lcccc@{}}
\toprule
\textbf{nDCG@10 }          & DL19  & DL20  & MS Dev & BEIR  \\ \midrule
Condenser      & 1.1  & 3.3  & 0.6   & 6.3  \\
+ \infoncesynth{}    & 58.1 & 57.0    & 28.3  & 37.1 \\
\textcolor{purple}{+ \wssynth{}}                & 63.3 & 55.9 & 29.7  & 39.3 \\ \midrule
CoCondenser-Marco & 31.1  & 33.7 & 14.0  & 31.0 \\
+ \infoncesynth{}    & 59.6 & 59.0 & 29.7  & 39.4  \\
\textcolor{purple}{+ \wssynth{}}                & 59.7 & 59.6 & 30.5  & 41.3 \\ \bottomrule
\end{tabular}
}
\caption{Self-supervised Condenser~\cite{Gao2021CondenserAP} and CoCondenser trained on MS~MARCO. The models are further fine-tuned on our synthetic data using InfoNCE with binarized labels or Wasserstein distance with the original 4-level labels.}
\label{tab:ccdenser_eval}
\end{table}

\section{Qualitative Examples}
\label{sec:qual_samples}

\paragraph{Sample Retrieved Passages}
\Cref{tab:retrieval_samples} shows the retrieved passages for a sample query by a model trained on binary ranking contexts with InfoNCE and another model trained on multi-level ranking contexts with Wasserstein distance.
Although both models identify the most relevant passage correctly, the model trained on multi-level ranking contexts has a better understanding of relevance and retrieves better passages in other ranks.

\begin{table*}[t]
\centering
\scriptsize
\setlength{\tabcolsep}{4pt}
\begin{tabular}{@{}p{0.32\textwidth}p{0.32\textwidth}p{0.32\textwidth}@{}} 

\multicolumn{3}{l}{\textbf{Query}: \textit{"border personality disorder symptoms"}} \\
\addlinespace[3pt]
\toprule
\makecell[c]{\textbf{Synthetic Documents}} & \makecell[c]{\textbf{E5 Mistral Candidates}} & \makecell[c]{\textbf{BM25 Candidates}} \\ \midrule

Borderline personality disorder (BPD) is a serious mental illness characterized by pervasive instability in moods, interpersonal relationships, self-image, and behavior. Symptoms of BPD include frantic efforts to avoid real or imagined abandonment, intense interpersonal relationships marked by alternating extremes of idealization and devaluation, and unstable self-image or sense of self. Individuals with BPD may also exhibit impulsive behaviors, such as excessive spending, reckless driving, or risky sex, and have recurring suicidal thoughts or self-mutilating behaviors.
&
The symptoms of borderline personality disorder include: a recurring pattern of instability in relationships, efforts to avoid abandonment, identity disturbance, impulsivity, emotional instability, and chronic feelings of emptiness, among other symptoms.
&
Description of Affective personality disorder Affective personality disorder: Related Topics These medical condition or symptom topics may be relevant to medical information for Affective personality disorder: Related Topics. These medical condition or symptom topics may be relevant to medical information for  Affective personality. Personality disorder (2 causes) Affective. Affective symptoms. Affective disorder. Personality.
\\ \midrule

Certain personality disorders, including borderline personality disorder, can have a significant impact on an individual's emotional and psychological well-being. People with these conditions may experience intense emotional dysregulation, leading to mood swings, irritability, and impulsive behaviors. They may also struggle with maintaining stable relationships, due to fear of abandonment or difficulty with emotional intimacy. While the exact causes of these disorders are not fully understood, treatment options such as dialectical behavior therapy and medication can help alleviate symptoms and improve overall functioning.
&
~ 5 min read. The symptoms of borderline personality disorder include: a recurring pattern of instability in relationships, efforts to avoid abandonment, identity disturbance, impulsivity, emotional instability, and chronic feelings of emptiness, among other symptoms.
&
Narcissistic Personality Disorder symptoms include a complete and total lack of empathy, along with a highly-exaggerated sense of self-importance.... narcissistic,personality,disorder,treatment,personality disorder treatment,narcissistic disorder symptoms,signs of narcissistic personality disorder,narcissistic personality disorder npd.
\\ \midrule

Emotional regulation is a critical aspect of mental health, and difficulties in this area can contribute to a range of psychological problems. Research has shown that individuals with mental health conditions, such as depression and anxiety, often struggle with managing their emotions in a healthy and adaptive way. This can lead to a range of negative consequences, including strained relationships, decreased productivity, and increased risk of self-destructive behaviors. By improving emotional regulation skills, individuals can better cope with stress and adversity, leading to improved overall well-being.
&
Borderline personality disorder (BPD) is a personality disorder that typically includes the following symptoms: 1  Inappropriate or extreme emotional reactions. 2  Highly impulsive behaviors. 3  A history of unstable relationships.
&
Personality disorder - Symptoms. Signs and symptoms of personality disorders. The different types of personality disorder that might need treatment can be broadly grouped into one of three clusters, called A, B or C. Cluster A personality disorders.
\\ \midrule

The city of Paris is known for its stunning architecture, rich history, and vibrant cultural scene. Visitors can explore famous landmarks like the Eiffel Tower, Notre-Dame Cathedral, and the Louvre Museum, which houses an impressive collection of art and artifacts from around the world. The city is also famous for its fashion industry, with top designers like Chanel and Dior showcasing their latest creations during Paris Fashion Week. Whether you're interested in history, art, or food, Paris has something to offer everyone.
&
By Mayo Clinic Staff. Borderline personality disorder affects how you feel about yourself, how you relate to others and how you behave. Signs and symptoms may include: An intense fear of abandonment, even going to extreme measures to avoid real or imagined separation or rejection.
&
Symptoms. Types of personality disorders are grouped into three clusters, based on similar characteristics and symptoms. Many people with one personality disorder also have signs and symptoms of at least one additional personality disorder.
\\ \bottomrule

\end{tabular}
\caption{Synthetic documents generated for one query compared to documents selected for pseudo labeling by BM25 and E5-Mistral-Instruct for the same query. The differences between synthetic documents are nuanced, and documents gradually change from perfectly relevant to irrelevant. All the selected candidates by E5-Mistral-Instruct are actually unannotated positives, and the candidates selected by BM25 are not actually relevant to the query or informative.}

\label{tab:pl_samples3}
\end{table*}

\begin{table*}[t]
\centering
\scriptsize
\setlength{\tabcolsep}{4pt}
\begin{tabular}{@{}lp{0.47\textwidth}p{0.47\textwidth}@{}} 

\multicolumn{3}{l}{\textbf{Query}: \textit{"how many ml a day should you drink"}} \\
\addlinespace[3pt]
\toprule
\textbf{rank} & \makecell[c]{\textbf{InfoNCE, binary relevance labels}} & \makecell[c]{\textbf{Wasserstein loss, 4-level relevance labels}} \\ \midrule

1
& 
\textcolor{ForestGreen}{How many ml of water should you drink in a day? The Institute of Medicine recommends that men drink 3000 ml of water each day and women drink 2100 ml. This equals approximately 13 cups of liquid for men and 9 cups from women. An individual often requires more water to stay hydrated in hot weather or due to strenuous exercise. <~continued~>.}
&
\textcolor{ForestGreen}{How many ml of water should you drink in a day? The Institute of Medicine recommends that men drink 3000 ml of water each day and women drink 2100 ml. This equals approximately 13 cups of liquid for men and 9 cups from women. An individual often requires more water to stay hydrated in hot weather or due to strenuous exercise.}	
\\ \midrule

2
& 
\textcolor{ForestGreen}{Since 2,000 mL of fluid are needed daily for normal body functions, first determine how many mL each patient has consumed so far today. Identify which patients need to be encouraged to consume more fluids to meet the 2,000 mL intake standard.}
&
\textcolor{ForestGreen}{How many ml of water should you drink in a day? The Institute of Medicine recommends that men drink 3000 ml of water each day and women drink 2100 ml. This equals approximately 13 cups of liquid for men and 9 cups from women. An individual often requires more water to stay hydrated in hot weather or due to strenuous exercise. <~continued~>}	
\\ \midrule

3
& 
\textcolor{ForestGreen}{How many ml of water should you drink in a day? The Institute of Medicine recommends that men drink 3000 ml of water each day and women drink 2100 ml. This equals approximately 13 cups of liquid for men and 9 cups from women. An individual often requires more water to stay hydrated in hot weather or due to strenuous exercise.}	
&
\textcolor{ForestGreen}{The recommendation stated that a person should have 1 ml (about 1/5 of a teaspoon) of water for each calorie he or she consumes. The average diet at the time was approximately 1900 calories, meaning you needed about 64 ounces of water per day. Now the Institute of Medicine sets general guidelines for total water intake.It recommends that women consume a total of 91 ounces (that's about 2.7 liters) per day from all food and beverages combined. For men, it's about 125 ounces a day (or 3.7 liters).Depending on your diet, about 25\% of the water you consume comes from your food. Most of us healthy folks get enough water in the foods and liquids we consume. That includes any liquid we drink  even caffeinated beverages like soda, coffee and tea.t recommends that women consume a total of 91 ounces (that's about 2.7 liters) per day from all food and beverages combined. For men, it's about 125 ounces a day (or 3.7 liters).}	
\\ \midrule

4
& 
\textcolor{SpringGreen}{How many quarts of water should you drink each day? The recommended minimum amount of water you should have each day is 8 cups, which is equal to 2 quarts.}	
&
\textcolor{ForestGreen}{The recommendation stated that a person should have 1 ml (about 1/5 of a teaspoon) of water for each calorie he or she consumes. The average diet at the time was approximately 1900 calories, meaning you needed about 64 ounces of water per day. Now the Institute of Medicine sets general guidelines for total water intake.It recommends that women consume a total of 91 ounces (that's about 2.7 liters) per day from all food and beverages combined.For men, it's about 125 ounces a day (or 3.7 liters).Depending on your diet, about 25\% of the water you consume comes from your food. Most of us healthy folks get enough water in the foods and liquids we consume. That includes any liquid we drink even caffeinated beverages like soda, coffee and tea.or men, it's about 125 ounces a day (or 3.7 liters). Depending on your diet, about 25\% of the water you consume comes from your food. Most of us healthy folks get enough water in the foods and liquids we consume. That includes any liquid we drink even caffeinated beverages like soda, coffee and tea.}	
\\ \midrule

5
& 
\textcolor{Maroon}{If you are overweight or workout vigorously, this number will increase. And then, if you want to lose weight, you can add 500 ml water to your regular water intake to burn around 23 calories per day that will help you lose upto 5 pounds of weight per year.}	
&
\textcolor{ForestGreen}{The Institute of Medicine recommends an average of 3.7 liters (125 ounces) per day for healthy adult men and 2.7 liters (91 ounces) per day for healthy adult women, allowing adjustments for activity and health levels, climate and elevation, and the amount of water consumed from food and other drinks.}	
\\ \midrule

6
& 
\textcolor{SpringGreen}{Presuming you're awake for approximately 16 hours per day, you'll have to drink between 4.65 and 6.25 fluid ounces per hour. That may seem like a lot, but it isn't much more than four to eight sips per hour (depending on how much you take in).}	
&
\textcolor{ForestGreen}{How much you would need to drink daily isn't clear, I would suggest just a bit more than 1 liter a day instead of the often quoted 2 - 3 liters a day. . Metabolic processes will generate about 300 ml of water a day, your food contains about 800 ml of water daily. The rest of your intake is what you drink.}	
\\ \midrule

7
& 
\textcolor{ForestGreen}{It means your normal urine output per hour should be anywhere between 33.3 and 83.3 ml. If it's not within this range, there's something wrong. However, you need to ensure that you're drinking no less than 2 liters of fluid per day. These numbers may change a bit considering your unique circumstances.}	
&
\textcolor{ForestGreen}{The Institute of Medicine advises that men consume roughly 3.0 liters (about 13 cups) of total beverages a day and women consume 2.2 liters (about 9 cups) of total beverages a day.}	
\\ \midrule

8
& 
\textcolor{Maroon}{However, each drinking session of three pints is at least six units, which is more than the safe limit advised for any one day. Another example: a 750 ml bottle of 12\% wine contains nine units. If you drink two bottles of 12\% wine over a week, that is 18 units.}	
&
\textcolor{SpringGreen}{How much should you drink: It's said we need 8-10 glasses of water a day (8 oz. glasses). That's at least 2 quarts of water. This is just to provide the water we need to wash away the acidity from our bodily functions and remove our own wastes.}	
\\ \midrule

9
& 
\textcolor{Maroon}{A 10 ml bottle contains 1000 units There are 100 units in a mL. 1 cc equals 100 units, so to figure how long a 10mL bottle, (1000 units) will last, you divide the number of units you use per day into 1000, and there you have it.Actually it depends on the concentration of the bag of solution you have. 10 ml bottle contains 1000 units There are 100 units in a mL. 1 cc equals 100 units, so to figure how long a 10mL bottle, (1000 units) will last, you divide the number of units you use per day into 1000, and there you have it.}	
&
\textcolor{SpringGreen}{How much water does one person need to drink per day? you should drink at least 7 to 10 average sized glasses of water each day. One average sized glass is about eight ounces. There are 16 ounces in a pint, 2 pints in a quart, and 4 quarts in a gallon, so, mathematically, there are about 128 ounces in a gallon.}	
\\ \midrule

10
& 
\textcolor{GreenYellow}{When I got to work, I filled up my 16-ounce water bottle and drank it through a straw. For some reason, drinking though a straw helped me to drink more because I would take sips without thinking. I had to fill this up six times per day to get the 3 liters. For the first few days, I made a conscious effort to keep up with this, but after day four, I started to write down how many ounces I drank just to keep track.}	
&
\textcolor{ForestGreen}{How many ml of water should you drink in a day? A: The Institute of Medicine recommends that men drink 3000 ml of water each day and women drink 2100 ml. This equals approximately 13 cups of liquid for men ... <~continued~>}	
\\ \midrule

\end{tabular}
\caption{Retrieved MS~MARCO (real) passages for a sample query by a Contriever trained on synthetic documents using binary labels with InfoNCE (left) and the same model trained on the same documents using multi-level ranking contexts with the Wasserstein distance as a loss, i.e. SyCL (right). SyCL trains models to distribute higher relevance scores over a larger number of documents. Here, only one of these documents is labeled as relevant in the dataset, although, in fact, many are \textcolor{ForestGreen}{relevant} or even near-duplicates; that makes them false negatives. This is a typical case in MS~MARCO, confounding training and evaluations that rely on these labels (as MM dev).}

\label{tab:retrieval_samples}
\end{table*}

\section{Additional Evaluation}

For our main experiments in \cref{sec:exp}, we also measure MRR@100 and Recall@100.
As shown in \cref{tab:main_mrr,tab:main_recall}, we observe similar improvements for SyCL compared to other methods.

\begin{table*}[t]
\centering
\setlength{\tabcolsep}{4pt}
\resizebox{\textwidth}{!}{
\begin{tabular}{@{}lccc|cccccc@{}}
\toprule
\textbf{MRR@100} & \textbf{DL19} & \textbf{DL20} & MM Dev & FEVER & HotpotQA & FiQA & NQ & Quora & Touche   \\
\cmidrule(lr){2-10}

Base Contriever (BC)                      & 76.0 & 78.8 & 17.4 & 64.3 & 64.0 & 31.0 & 23.1 & 82.6 & 38.6 \\ \midrule
BC + \infoncesynth{}                      & 84.0 & 76.9 & 22.6 & 65.6 & 61.9 & 34.2 & 29.8 & 74.8 & 31.8 \\
\textcolor{purple}{BC + \wssynth{}}       & 93.8 & 90.5 & 26.0 & 82.6 & 76.1 & 35.0 & 37.9 & 82.6 & 41.8 \\ \cdashlinelr{1-10}
BC + \infoncereal{}                       & 91.3 & 87.1 & 29.5 & 67.9 & 78.1 & 36.2 & 38.4 & 80.6 & 30.1 \\
\textcolor{purple}{BC + \wswithreal{}}    & 92.3 & 87.7 & 28.1 & 81.2 & 78.7 & 37.4 & 38.1 & 82.9 & 36.9 \\

\end{tabular}
}
\resizebox{\textwidth}{!}{
\begin{tabular}{@{}lccccccccc@{}}
\toprule
\textbf{MRR@100} & \makecell{CQADup\\Android} & Scidocs & \makecell{Climate\\FEVER} & DBPedia & \makecell{TREC\\COVID} & Scifact & NFCorpus & ArguAna & \makecell{\textbf{BEIR}\\\textbf{Avg}} \\ \cmidrule(l){2-10} 

Base Contriever (BC)                      & 38.3 & 29.0 & 21.3 & 59.9 & 58.0 & 60.2 & 51.6 & 21.6 & 46.0 \\ \midrule
BC + \infoncesynth{}                      & 36.2 & 28.5 & 29.5 & 63.5 & 49.2 & 59.2 & 51.7 & 18.5 & 45.3 \\
\textcolor{purple}{BC + \wssynth{}}       & 39.1 & 31.2 & 37.9 & 73.8 & 73.5 & 58.5 & 52.5 & 19.5 & 53.0 \\ \cdashlinelr{1-10}
BC + \infoncereal{}                       & 38.7 & 29.8 & 26.3 & 70.8 & 57.2 & 62.4 & 51.3 & 23.3 & 49.4 \\
\textcolor{purple}{BC + \wswithreal{}}    & 40.7 & 29.8 & 35.5 & 75.0 & 74.3 & 64.3 & 53.2 & 22.5 & 53.6 \\

\bottomrule

\end{tabular}
}
\caption{Retrieval effectiveness (MRR@100). Base Contriever (BC): self-supervised Contriever model. `BC +' denotes the fine-tuning setting in terms of \textbf{loss function}: InfoNCE / Wasserstein (WS), and \textbf{type of data}: real data from the MS~MARCO training set with annotated positives and mined hard negatives (Real) / fully synthetic multi-level documents (Synth.) / combination. DL19, DL20, and MM Dev are the TREC DL 2019, TREC DL 2020, and Dev evaluation sets of MS~MARCO. Evaluation on the rest of sets is zero-shot. \textcolor{purple}{Purple: SyCL, our method.}}
\label{tab:main_mrr}
\end{table*}

\begin{table*}[t]
\centering
\setlength{\tabcolsep}{4pt}
\resizebox{\textwidth}{!}{
\begin{tabular}{@{}lccc|cccccc@{}}
\toprule
\textbf{Recall@100} & \textbf{DL19} & \textbf{DL20} & MM Dev & FEVER & HotpotQA & FiQA & NQ & Quora & Touche   \\
\cmidrule(lr){2-10}

Base Contriever (BC)                      & 41.8 & 44.6 & 67.2 & 93.3 & 70.5 & 58.0 & 77.2 & 98.7 & 41.9 \\ \midrule
BC + \infoncesynth{}                      & 44.0 & 47.9 & 74.1 & 93.8 & 66.7 & 60.0 & 83.1 & 97.5 & 39.7 \\
\textcolor{purple}{BC + \wssynth{}}       & 44.7 & 49.4 & 77.6 & 95.3 & 71.9 & 60.0 & 86.7 & 98.8 & 46.3 \\ \cdashlinelr{1-10}
BC + \infoncereal{}                       & 48.3 & 53.1 & 84.1 & 93.3 & 75.7 & 63.7 & 90.0 & 98.8 & 41.8 \\
\textcolor{purple}{BC + \wswithreal{}}    & 49.0 & 54.6 & 82.8 & 95.1 & 74.8 & 64.3 & 89.5 & 98.9 & 44.0 \\

\end{tabular}
}
\resizebox{\textwidth}{!}{
\begin{tabular}{@{}lccccccccc@{}}
\toprule
\textbf{Recall@100} & \makecell{CQADup\\Android} & Scidocs & \makecell{Climate\\FEVER} & DBPedia & \makecell{TREC\\COVID} & Scifact & NFCorpus & ArguAna & \makecell{\textbf{BEIR}\\\textbf{Avg}} \\ \cmidrule(l){2-10} 

Base Contriever (BC)                      & 74.5 & 36.0 & 45.6 & 45.3 & 3.7 & 90.4 & 29.3 & 94.7 & 61.4 \\ \midrule
BC + \infoncesynth{}                      & 72.1 & 35.5 & 51.5 & 45.0 & 3.3 & 92.2 & 29.2 & 89.5 & 61.4 \\
\textcolor{purple}{BC + \wssynth{}}       & 76.8 & 36.1 & 57.3 & 46.8 & 8.8 & 92.8 & 30.5 & 94.0 & 64.4 \\ \cdashlinelr{1-10}
BC + \infoncereal{}                       & 72.9 & 36.6 & 45.3 & 49.9 & 3.8 & 91.1 & 29.9 & 96.2 & 63.5 \\
\textcolor{purple}{BC + \wswithreal{}}    & 75.5 & 36.5 & 56.0 & 50.8 & 8.4 & 93.3 & 31.1 & 97.1 & 65.4 \\

\bottomrule

\end{tabular}
}
\caption{Retrieval effectiveness (Recall@100). Base Contriever (BC): self-supervised Contriever model. `BC +' denotes the fine-tuning setting in terms of \textbf{loss function}: InfoNCE / Wasserstein (WS), and \textbf{type of data}: real data from the MS~MARCO training set with annotated positives and mined hard negatives (Real) / fully synthetic multi-level documents (Synth.) / combination. DL19, DL20, and MM Dev are the TREC DL 2019, TREC DL 2020, and Dev evaluation sets of MS~MARCO. Evaluation on the rest of sets is zero-shot. \textcolor{purple}{Purple: SyCL, our method.}}
\label{tab:main_recall}
\end{table*}

\section{Prompting Details}
\label{sec:app_prompting}

\Cref{tab:full_prompt} shows the exact prompt that we used to generate multi-level ranking contexts for training queries of the MS-MARCO dataset.
To create in-context examples, we use the annotations in the TREC DL 2023 split.
For each prompt, we randomly sample one query and four passages (one for each relevance level in TREC DL 2023 annotations) and use them as the in-context example.
To increase the diversity of the generated passages, for each prompt, we randomly sample the value of \rinlinecode{\{\{num\_sentences\}\}} from \rinlinecode{\{none,2,5,10,15\}} with probabilities \rinlinecode{\{0.5,0.1,0.2,0.1,0.1\}}.
Similarly, we randomly sample the value of \rinlinecode{\{\{difficulty\_level\}\}} from \rinlinecode{\{none, high school, college, PhD\}} with probabilities \rinlinecode{\{0.4,0.2,0.2,0.2\}}.
For both variables, if the sampled value is \rinlinecode{none}, we do not include the corresponding instruction in the prompt.

We also noticed that the LLM has a tendency to provide the exact answer to the query in the very first sentence of the perfectly relevant passage.
To avoid such spurious patterns, in 30\% of the prompts, we include an additional instruction and explicitly ask the LLM to avoid answering the query in the very first sentence of the perfectly relevant passage.

\paragraph{Direct Binary Data Generation}

In \cref{sec:additional_analysis}, we adapt the short-long matching prompt in Table 8 of \citet{wang-etal-2024-improving-text} to generate one positive and two negatives for existing queries, which matches the ranking context size in our experiments. Specifically, we use the prompt in \cref{tab:direct_bin_prompt} to directly generate these binary passages.

\section{Implementation Details}
\label{sec:imp_details}
We train our models for only one epoch using the Trainer module in the Huggingface transformers library\footnote{https://github.com/huggingface/transformers}.
For both training and evaluation, we use the maximum length of 256 for both queries and passages.
We use a total batch size of 64 across four GPUs (batch size of 16 per device).
We set the learning rate to 1e-5, gradient accumulation steps to 4, and warm-up ratio to 0.05.
We use the default parameters in version 4.48.0 of the transformers library for all other configurations, e.g., optimizer, learning rate scheduler, etc.
Each one of our experiments takes about 2 hours using one machine with four L40s GPUs.

Note that since the original Contriever paper~\citep{izacard2021contriever} uses a sequence length of 512, our evaluation results are slightly different from what is reported in \citet{izacard2021contriever}.

\paragraph{Data Generation Costs}
We have generated our data locally over many sessions using different GPU devices, which unfortunately, makes calculating exact cost figures challenging.
Here, we approximate the costs based on the number of tokens used for generating data for the 502,000 training queries in MS~MARCO.
Since our data is very similar to MS~MARCO, we use an average length of 128 and 32 tokens for each passage and query, respectively.
This is an overestimation, and the actual average length of each passage and query in MS~MARCO is 80 and 10 tokens, respectively.
For a reasonable approximation, we use the prices of GPT-4o Mini batch API at the time of writing (input: \$0.075/1M, output: \$0.30/1M), which leads to \textasciitilde \$100 for the cost of API calls.

Note that we use public models that can be deployed on local hardware, which reduces costs.
More importantly, we show that we can generate data of comparable quality with smaller 32B LLMs.
Inference with a 32B model is drastically cheaper, which makes synthetic data generation even more appealing.

\section{Other Models}
\label{sec:other_models}
We repeat our main experiments using Condenser~\citep{Gao2021CondenserAP} and CoCondenser-Marco~\citep{gao2021unsupervised} as the base retrievers.
Condenser is a BERT model with a slight architectural modification during pre-training that makes the learned representations more suitable for retrieval.
CoCondenser-Marco is a Condenser model fine-tuned on the MS~MARCO corpus in an \textit{unsupervised manner} (i.e., without using any labels).
Since these two models do not perform as well as the Contriever model, we train them for three epochs instead of one and also increase the learning rate to 1e-4.
As shown in~\cref{tab:ccdenser_eval}, synthetic data significantly improves the base unsupervised model in both cases.
Moreover, except for Condenser on the DL20 split of MS MARCO, training using multiple relevance labels leads to better performance compared to contrastive training with binary labels using the InfoNCE loss.
Notably, the base Condenser model is only trained with a language modeling objective without any retrieval-specific fine-tuning, which could potentially impact its ability to learn the nuanced differences between multiple levels of relevance.
Furthermore, we noticed that Wasserstein loss leads to smaller gradient norms than InfoNCE loss (i.e., smaller updates and thus slower convergence).
As a result, we speculate that for lower-quality models or models without contrastive pre-training, the difference between InfoNCE and Wasserstein losses will increase with more training steps.

\section{Loss Functions}
\label{sec:apndx_losses}

We calculate the similarity between query $q$ and document $d$ as the inner product between their embeddings. Specifically,
\begin{equation*}
\operatorname{sim}(d, q) = f_\theta(d) \cdot f_\theta(q) \, ,
\end{equation*}
where $f$ is the embedding function parameterized by $\theta$.

\paragraph{InfoNCE}
We calculate the InfoNCE loss as follows:

\begin{equation*}
    -\log \frac{\operatorname{exp}(\operatorname{sim}(d^+, q))}{\sum_{d \in D_q} \operatorname{exp}(\operatorname{sim}(d, q))} \, ,
\end{equation*}
where $d^+$ is the positive document, and $D_q$ is the ranking context for query $q$ (i.e., the collection of positive and negative documents for $q$).
Note that for InfoNCE loss, $D_q$ can contain one and only one positive document, and the rest must be negative.

\paragraph{KL Divergence}
Given the similarity scores between a query and documents in its ranking context, we calculate the KL loss as follows:
\begin{equation*}
    \operatorname{D_{KL}}(\sigma (Y) \| \sigma (\hat{Y})) \, ,
\end{equation*}
where $\sigma$ is the softmax function, and $Y \in \mathbb{R}^{\left|D_q\right|}$ and $\hat{Y} \in \mathbb{R}^{\left|D_q\right|}$ are the ground truth relevance labels and predicted relevance labels (i.e., similarity scores) for documents in the ranking context of query $q$, respectively.

\paragraph{Wasserstein Distance}
We use the special case of Wasserstein distance between two multivariate Gaussian distributed inputs $X \sim \mathcal{N}(\mu_x, C_x)$ and $Y \sim \mathcal{N}(\mu_y, C_y)$, where $\mu$ and $C$ are the mean and covariance of each distribution, respectively.
For Gaussian distributions, the 2-Wasserstein distance reduces to
\begin{equation*}
   \operatorname{D}(X, Y) = \|\mu_x - \mu_y\|^2 - \operatorname{tr}(C_x + C_y - 2(C_xC_y)^{\frac{1}{2}}) \, .
\end{equation*}

In our implementation, we calculate the Wasserstein score for the entire batch.
Specifically, for each batch, we create matrices $H \in \mathbb{R}^{b\times\left|D_q\right|}$ and $\hat{H} \in \mathbb{R}^{b\times\left|D_q\right|}$ of shape \texttt{(batch size, ranking context size)} and minimize $\operatorname{D}(H, \hat{H})$ during training.
Each row of $H$ corresponds to ground truth relevance labels for one query in the batch.
Similarly, one row of $\hat{H}$ corresponds to the predicted similarity scores between one query in the batch and documents in its ranking context.
We use the fast implementation\footnote{\url{https://gist.github.com/Flunzmas/6e359b118b0730ab403753dcc2a447df}} proposed by~\citet{mathiasen2020backpropagating}.

\begin{table*}[t]
\footnotesize
\centering
\begin{tabular}{|p{0.99\linewidth}|}

\hline

\tt
\# Task\newline

You have been assigned a user query. Your mission is to write one positive passage and two negative passages for the given query.\newline
- "Positive Passage" is a relevant passage for the user query.\newline
- "Negative Passage" is a passage that only appears relevant to the query.\newline

Please adhere to the following guidelines:\newline
- All passages must be created independent of the query. Avoid copying the query verbatim. It's acceptable if some parts of the "Positive Passage" are not topically related to the query.\newline
- All passages should be at least {num\_sentences} sentences long.\newline
- The "Negative Passage" contains some useful information, but it should be less useful or comprehensive compared to the "Positive Passage".\newline
- Do not provide any explanation in any passages on why it is relevant or not relevant to the query.\newline
- The passages require {difficulty\_level} level education to understand.\newline

Do not explain yourself or output anything else. Be creative!
\\
\hline

\end{tabular}
\caption{Our prompt for directly generating binary passages for each query.}
\label{tab:direct_bin_prompt}
\end{table*}
\begin{table*}[t]
\footnotesize
\centering
\begin{tabular}{|l|p{0.87\linewidth}|}

\hline

Type & Content \\
\hline

System &
\tt \# Task\newline
\newline
You are a data engineer whose goal is to generate synthetic passages that teach a ranking system to sort a collection of passages based on how relevant they are to the user's search query (similar to a web search engine). Given a text query, your mission is to write four different passages, each with a different level of relevance to the given query. Specifically, you should write one passage for each of the following relevancy levels:\newline
- "Perfectly relevant passage": a passage that is dedicated to the query and contains the exact answer.\newline
- "Highly relevant passage": a passage that has some answer for the query, but the answer may be a bit unclear, or hidden amongst extraneous information.\newline
- "Related passage": a passage that seems related to the query but does not answer it.\newline
- "Irrelevant passage": a passage that has nothing to do with the query.\newline
\newline
\#\# Passage generation instructions\newline
- All passages should be about \{\{num\_sentences\}\} sentences long.\newline
- All passages require \{\{difficulty\_level\}\} level education to understand.\newline
- \{\{The very first sentence of the passage must NOT completely answer the query.\}\}\newline
- Avoid copying the query verbatim. It’s acceptable if some parts of the "Perfectly relevant passage" are not topically related to the query.\newline
- How related each passage is to the given query should closely adhere to the corresponding relevancy level.\newline
- Passages can be less relevant to a given query for different reasons. For example, they might be less useful, less accurate, less comprehensive, etc. Explore different ways for writing less relevant passages. Be creative!\newline
- Do not provide any explanation in any passage on why it is relevant or not relevant to the query.\newline
\newline
\#\# Evaluation criteria\newline
To double check if you have successfully accomplished the task, you should imagine how a search engine like Google Search would rank the generated passages if you search for the given query. To accomplish the task successfully, a search engine like Google Search should rank your passages in the same order that you generated them. In other words:\newline
- the perfectly relevant passage should fully answers the query.\newline
- the highly relevant passage should be less relevant to the query than the perfectly relevant passage.\newline
- the related passage should be less relevant to the query than the highly relevant passage.\newline
- the irrelevant passage should not provide any useful information about the query.\newline
\newline
Do not explain yourself or output anything else. Be creative!
\\

\hline

User &
\tt \#\# Query: \{\{IC\_example\_query\}\}
\\
\hline 

Assistant&
\tt [Perfectly relevant passage]\newline
\newline
\{\{IC\_example\_perfectly\_relevant\_passage\}\}\newline
\newline
[Highly relevant passage]\newline
\newline
\{\{IC\_example\_highly\_relevant\_passage\}\}\newline
\newline
[Related passage]\newline
\newline
\{\{IC\_example\_related\_passage\}\}\newline
\newline
[Irrelevant passage]\newline
\newline
\{\{IC\_example\_irrelevant\_passage\}\}
\\
\hline

User
&
\tt \#\# Query: \{\{main\_query\}\}
\\
\hline

\end{tabular}
\caption{Our full prompt template used to generate synthetic multi-level ranking contexts for each query. See~\cref{sec:app_prompting} for more details.}
\label{tab:full_prompt}
\end{table*}

\end{document}